\documentclass[onecolumn,prd,amsmath,amssymb,superscriptaddress,nofootinbib,10pt]{revtex4-1}
\usepackage{color}
\pdfoutput=1
\usepackage{graphicx}
\usepackage[latin1]{inputenc}
\usepackage{hyperref}
\usepackage{epstopdf}
\usepackage{slashed}
\usepackage{cancel}
\usepackage{epsfig,psfrag,graphics,verbatim}
\usepackage{dcolumn}% Align table columns on decimal point
\usepackage{bm}% bold math
\usepackage{slashed}
\usepackage[]{units}
\usepackage{ulem}
\usepackage{url}
\usepackage{array}
\usepackage{makecell}
\usepackage{soul}
\usepackage{tikz}
\usepackage{forest}

 \def\be   {\begin{equation}}  
  \def\ee   {\end{equation}}
 \def\ba   {\begin{array}}     
  \def\ea   {\end{array}}
 \def\bea  {\begin{eqnarray}}  
  \def\eea  {\end{eqnarray}}
 \def\bean {\begin{eqnarray*}}  
 \def\eean {\end{eqnarray*}}

\definecolor{darkgreen}{rgb}{0,0.5,0}

\newcommand{\AddrSDU}{
CP$^{3}$-Origins, University of Southern Denmark, Campusvej 55, DK-5230 Odense M, Denmark
}

\begin{document}

{\small
\begin{flushright}
CP3-Origins-2018-043 DNRF90
\end{flushright} }

\mathversion{bold}
\title{Lepton and Quark Mixing from Stepwise Breaking of Flavor and CP}
 \mathversion{normal}
 
 \author{Claudia Hagedorn}\email{hagedorn@cp3.sdu.dk}
 \affiliation{\AddrSDU}

\author{Johannes K\"{o}nig}\email{konig@cp3.sdu.dk}
 \affiliation{\AddrSDU}
 
\begin{abstract}

We explain all features of lepton and quark mixing in a scenario with the flavor symmetry $\Delta (384)$ and a CP symmetry, 
 where these are broken in several steps. The residual symmetry in the neutrino and up quark
sector is a Klein group and CP, while a $Z_3$ and a $Z_{16}$ symmetry are preserved
 among charged leptons and down quarks, respectively. If the Klein group in the
 neutrino sector is further broken down to a single $Z_2$ symmetry, we obtain 
 predictions for all lepton mixing parameters in terms of one real quantity, whose
  size is determined by the value of the reactor mixing angle. The Dirac and Majorana
 phases are fixed, in particular $\sin\delta \approx -0.936$. 
   A sum rule, relating these CP phases and the reactor and atmospheric mixing angles, $\theta_{13}$ and $\theta_{23}$, is given. 
  In the quark sector, we have for the Cabibbo angle $\theta_C= \sin \pi/16 \approx 0.195$
  after the first step of symmetry breaking. This is brought into full accordance with experimental data with the second step of symmetry breaking, where either the $Z_{16}$ group
 is broken to a $Z_8$ symmetry in the down quark sector or the Klein group to one $Z_2$ symmetry only among up quarks. 
 The other two quark mixing angles are generated in the third and last symmetry breaking step, when the residual symmetries in the up and/or down quark sector are further broken.
  If this step occurs among both up and down quarks, the amount of CP violation in the quark sector 
  is determined by the lepton sector and explaining the current neutrino oscillation data entails that the Jarlskog invariant 
  $J_{\mbox{\tiny CP}}^q$ is in very good agreement with experimental findings.
   Lastly, a sum rule is derived that contains the CP phase $\delta^q$ and $\theta_C$ of the quark sector and the lepton mixing parameters $\theta_{13}$, $\theta_{23}$ and $\delta$.

\end{abstract}

\maketitle

%%%%%%%%%%%%%%%%%%%%%%%%%%%%%%%%%%%%%
\section{Introduction}
\label{sec:Introduction}
%%%%%%%%%%%%%%%%%%%%%%%%%%%%%%%%%%%%%

Most of the free parameters in the Standard Model (SM) of particle physics are related to fermion masses and mixing.
In particular, the size of the mixing angles in the quark and lepton sectors and their striking difference as well as the amount of CP violation remain unexplained.
 Approaches with discrete, non-abelian flavor symmetries $G_f$ have been successfully employed in order to determine the mixing angles of 
 leptons and quarks, for reviews on flavor symmetries see~\cite{Altarelli:2010gt,Ishimori:2010au,King:2013eh,Grimus:2011fk}. 
  Those, where different residual symmetries remain preserved in the charged lepton (down quark) and neutrino (up quark) sectors, can determine
 all mixing angles and the Dirac-type CP phase~\cite{Lam:2007qc,Blum:2007jz,Lam:2008rs}. Even more predictive are approaches with a flavor and a CP symmetry~\cite{Harrison:2002kp,Grimus:2003yn,Feruglio:2012cw}, since these
allow to fix all CP phases with the help of the choice of the residual symmetries. Many of the analyses focus on
the lepton sector~\cite{Lam:2008rs,deAdelhartToorop:2011re,King:2013vna,Hagedorn:2013nra,Fonseca:2014koa,Feruglio:2012cw,Hagedorn:2014wha,Ding:2014ora,Talbert:2014bda,Everett:2015oka} and only rather few attempt to (also) determine the mixing in the quark sector by the mismatch of residual symmetries in up and down quark sectors~\cite{Lam:2007qc,Blum:2007jz,Holthausen:2013vba,Araki:2013rkf,Yao:2015dwa,Varzielas:2016zuo,Li:2017abz,Lu:2018oxc}.
 
We consider as $G_f$ the discrete group $\Delta (384)$ and a CP symmetry that corresponds 
to an automorphism of $G_f$~\cite{Grimus:1995zi,Feruglio:2012cw,Holthausen:2012dk,Chen:2014tpa}. 
Being primarily interested in fermion mixing in this study, we focus on the three generations of left-handed (LH) lepton doublets $L_i$ and quark doublets $Q_i$, $i=1,2,3$, 
not specifying the transformation properties of the right-handed (RH) fields. Furthermore, we assume that neutrinos are Majorana particles, whose mass is generated from the Weinberg
operator. The three generations of LH lepton doublets and quark doublets both transform as a faithful, irreducible, complex three-dimensional representation ${\bf 3}$
of $\Delta (384)$. The symmetries $G_f$ and CP are broken by an unspecified mechanism to different residual symmetries in the neutrino and up quark sector,
charged lepton as well as down quark sector. This first step of symmetry breaking leads to tri-bimaximal (TB) mixing~\cite{Harrison:2002er,Xing:2002sw} in the lepton sector and generates the
Cabibbo angle $\theta_C=\sin\pi/16 \approx 0.195$ in the quark sector~\cite{Blum:2007jz,deAdelhartToorop:2011re}. 
 In a further step of symmetry breaking, where the residual symmetry among neutrinos is reduced, 
the reactor mixing angle is induced and all lepton mixing parameters become functions of one real quantity, whose
size is determined by the reactor mixing angle $\theta_{13}$. Accommodating the measured value of the atmospheric mixing angle $\theta_{23}$ well 
 and the indication for close to maximal CP violation in neutrino oscillations~\cite{Esteban:2016qun}, one CP symmetry is singled out that leads to $\sin^2 \theta_{23} \approx 0.579$ 
 and $\sin \delta \approx -0.936$. 
  For this choice of CP symmetry the Majorana phases $\alpha$ and $\beta$ have to fulfil $|\sin\alpha|=|\sin\beta| =1/\sqrt{2}$. 
  Both Majorana phases fulfil a non-trivial relation, involving the CP phase $\delta$ and the two mixing angles $\theta_{13}$ and $\theta_{23}$.
   The quantity $m_{ee}$, 
 measurable in neutrinoless double beta decay, becomes strongly constrained, e.g.~for positive $\sin\alpha$ and $\sin\beta$ we find $m_{ee} \gtrsim 2.86 \, \times \, 10^{-3} \, \mathrm{eV}$ 
  for neutrino masses with normal ordering and for inverted ordering that $m_{ee}$ is detectable with the next generation of experiments~\cite{DellOro:2016tmg}. 
 In the quark sector, either the residual group in the up or the down quark sector is reduced at the second step of symmetry breaking, so that 
 the Cabibbo angle is brought into full accordance with experimental data~\cite{PDG2018}. Breaking the residual
symmetries among up and/or down quarks even further eventually gives rise to the remaining two quark mixing angles. 
 If this breaking occurs in only one of the two sectors, the amount of CP violation in quark mixing will depend on phases, that are in general not specified further by the residual symmetries. If instead
 the latter are broken in the down as well as the up quark sector, a strong correlation between CP violation in the quark and in the lepton sector can be established and the Jarlskog
 invariant $J_{\mbox{\tiny CP}}^q$~\cite{Jarlskog:1985ht} can be determined to be $J_{\mbox{\tiny CP}}^q \approx 3.29 \times 10^{-5}$, in very good agreement with experimental data~\cite{PDG2018}.
  This strong correlation leads to a sum rule, relating the CP phase $\delta^q$ and the Cabibbo angle $\theta_C$ in the quark sector to the lepton mixing parameters $\delta$, $\theta_{13}$ and $\theta_{23}$. 
   These symmetry breaking sequences differ in several aspects from the symmetry breaking pattern that has been proposed in~\cite{Li:2017abz,Lu:2018oxc} in order to describe lepton and quark mixing 
   with the help of a flavor group $\Delta (6 \, n^2)$, $n$ integer, and CP.
  
The remainder of the paper is organized as follows: in section~\ref{sec:framework} we present basic information about $\Delta (384)$ and the employed CP symmetry. 
 In sections~\ref{sec:leptons}
and ~\ref{sec:quarks} we show the results for fermion mixing, arising from the different steps of symmetry breaking, in the lepton and quark sector, respectively. We summarize
and conclude in section~\ref{sec:conclusions}.
 
 %%%%%%%%%%%%%%%%%%%%%%%%%%%%%%%%%%%%% 
 \mathversion{bold}
\section{Basics about $\Delta (384)$ and CP}
\mathversion{normal}
\label{sec:framework}
%%%%%%%%%%%%%%%%%%%%%%%%%%%%%%%%%%%%%

We use as $G_f$ the group $\Delta (384)$, which belongs to the series $\Delta (6 \, n^2)$, $n$ integer. 
This group can be written as $(Z_8 \times Z_8) \rtimes S_3$ with $S_3$ being the permutation group of three distinct objects. 
 It can be described with four generators, $a$, $b$, $c$ and $d$, that fulfil the relations
\begin{eqnarray}
\nonumber
&&a^3=e \; , \;\; b^2=e \; , \;\; c^8 =e \; , \;\; d^8 =e \; ,
\\
\label{eq:generators}
&&(a \, b)^2= e \; , \;\; c \, d =d \, c\; , \;\; a \, c \, a^{-1} = c^{-1} d^{-1} \; , \;\; a \, d \, a^{-1} = c \; ,  \;\; b \, c \, b^{-1} = d^{-1} \; , \;\; b \, d \, b^{-1} = c^{-1}
\end{eqnarray} 
 with $e$ being the neutral element of the group~\cite{Escobar:2008vc}.
  The group $\Delta (384)$ has several subgroups, among them Klein groups $Z_2 \times Z_2$, $Z_3$ and $Z_{16}$ symmetries~\cite{deAdelhartToorop:2011re}.
 It possesses in total 13 Klein groups: twelve of these are conjugate to each other and can be generated by $c^4$ and $a \, b \, c^k$, $d^4$ and $a^2 \, b \, d^k$ and $(c \, d)^4$ and 
$b \, c^k \, d^k$ for $k=0,1,2,3$, respectively, while the remaining one is normal and can be obtained from $c^4$ and $d^4$ as generators. 
There are 64 $Z_3$ symmetries contained in $\Delta (384)$ and these can be described by $a \, c^i \, d^j$ 
 for $0 \leq i,j \leq 7$.
The twelve $Z_{16}$ subgroups can be generated by $b \, d^{2 \, l +1}$, $a \, b \, d^{2 \, l +1}$ and $a^2 \, b \, c^{2 \, l +1}$ for $l=0,1,2,3$.  All $Z_3$ and $Z_{16}$ groups are conjugate to each other, respectively.
For a comprehensive list of subgroups and their properties see~\cite{deAdelhartToorop:2011re,Ding:2014ora}.

As is known, CP symmetries are automorphisms of $G_f$~\cite{Grimus:1995zi,Feruglio:2012cw,Holthausen:2012dk,Chen:2014tpa} and a large class of these have been discussed in \cite{Hagedorn:2014wha} for the groups $\Delta (6 \, n^2)$. In the present study we consider 
the ones corresponding to the automorphisms composed by 
\begin{equation}
\label{eq:automorphism}
a \;\; \rightarrow \;\; a \;\; , \;\;  b \;\; \rightarrow \;\; b  \;\; , \;\;  c \;\; \rightarrow \;\; c^{-1} \;\; \mbox{and} \;\; d \;\; \rightarrow \;\; d^{-1}
\end{equation}
and the group transformation $a \, b \, c^s \, d^{2 \, s}$, $0 \leq s \leq 7$. 

The three generations of LH lepton doublets $L_i$ and quark doublets $Q_i$, $i=1,2,3$, are assigned to the same faithful, irreducible, complex three-dimensional representation ${\bf 3}$, 
which can be represented by the generators $a ({\bf 3})$, $b ({\bf 3})$, $c ({\bf 3})$ 
\begin{equation}
\label{eq:gens3}
a ({\bf 3}) =  \left( \begin{array}{ccc}
0 & 1 & 0\\
0 & 0 & 1\\
1 & 0 & 0
\end{array}
\right)
\;\; , \;\;
b ({\bf 3}) =  \left( \begin{array}{ccc}
0 & 0 & 1\\
0 & 1 & 0\\
1 & 0 & 0
\end{array}
\right)
\;\; , \; 
c({\bf 3})=  \left( \begin{array}{ccc}
\omega_8 & 0 & 0\\
0 & \omega_8^7 & 0\\
0 & 0 & 1
\end{array}
\right)
\end{equation}
and $d ({\bf 3})=a({\bf 3})^2 c ({\bf 3}) a ({\bf 3})$ with $\omega_8= e^{2 \pi i/8}$.\footnote{We could have chosen any of the other faithful, irreducible, complex three-dimensional representations
of $\Delta (384)$. Note that this can lead to the residual symmetries having different generators in terms of the elements of the flavor group in order to achieve the same results for mixing angles and 
  CP phases, compare also~\cite{deAdelhartToorop:2011re,King:2013vna}.} 
In the representation ${\bf 3}$ the mentioned type of CP symmetry corresponds to the CP transformation $X ({\bf 3})$ 
\begin{equation}
\label{eq:CP3}
X ({\bf 3}) (s) = a  ({\bf 3}) \, b  ({\bf 3})\, c ({\bf 3})^s \, d  ({\bf 3})^{2 \, s} \, X_0  ({\bf 3}) \; , \;\; 0 \leq s \leq 7
\end{equation}
with $X_0  ({\bf 3})$ representing the CP symmetry induced by the automorphism in Eq.~(\ref{eq:automorphism}) and being of the form of the identity matrix in the used basis.

Residual symmetries in the different fermion sectors of the theory are abelian subgroups of $G_f$, possibly together with the CP symmetry. From their mismatch
fermion mixing arises. 
  In particular, quark mixing is due to the mismatch of the residual group $G_u$ in the up quark and the one, called $G_d$, in the down quark sector, while 
 lepton mixing comes from the mismatch of $G_\nu$ and $G_l$, the residual symmetries in the neutrino and charged lepton sector, respectively.

\begin{figure}[t!]
\begin{tikzpicture}[level distance=1.8cm,
level 1/.style={sibling distance=6cm},
level 2/.style={sibling distance=3cm}]
\tikzstyle{every node}=[rectangle,rounded corners,draw]

\node (Root) {$\Delta (384)$ and CP}
    child {
    node {$G_l=Z_3$}  edge from parent node[right,draw=none] {$\phantom{xxxxx}$ TB mixing} 
}
child {
    node {$G_{\nu, 1}= G_{u, 1}= Z_2 \times Z_2 \times CP$} 
    child { node {$G_{\nu, 2}=Z_2 \times CP$} edge from parent node[left,draw=none] {$\theta_{13} \neq 0 \phantom{x}$}}
    child { node {$G_{u, 2}=Z_2 \times CP$} 
       child { node {$CP$} edge from parent node[right,draw=none] {$\phantom{x} \theta_{23}^q \neq 0$}
       } edge from parent node[right,draw=none] {$\phantom{x} |V_{us}| = 0.22452$}
    }
}
child{
  node {$G_{d,1}^\pm=Z_{16}$} 
   child { node[yshift=-1.8cm] {no residual} edge from parent node[left,draw=none,yshift=-0.8cm] {$\theta_{13}^q \neq 0 \phantom{x}$}
  } edge from parent node[left,draw=none] {$\theta_C=\sin\pi/16 \phantom{xxxx}$} 
};
\end{tikzpicture}
\caption{{\bf Stepwise breaking of \mathversion{bold}$\Delta (384)$\mathversion{normal} and CP} The stepwise breaking of the flavor and CP symmetry in the different sectors of the theory
 and the corresponding results for lepton and quark mixing 
 are shown. There are two steps of symmetry breaking in the lepton sector, while three steps are necessary in the quark sector in order to generate all three
quark mixing angles. The particular choice of the residual symmetries in the second and third step in the quark sector corresponds to the minimal viable case that leads to a strong correlation
between the amount of CP violation in the lepton and in the quark sector. The results for quark mixing angles and the Jarlskog invariant $J_{\mbox{\tiny CP}}^q$ can be found in section~\ref{sec:quarks}, 
while those for the lepton mixing parameters are detailed in section~\ref{sec:leptons}.
\label{fig1}}
\end{figure}
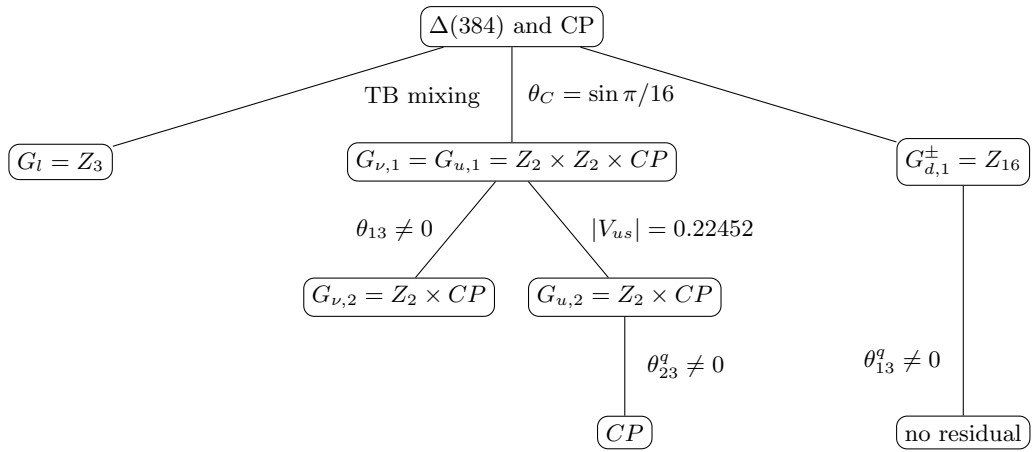

 %%%%%%%%%%%%%%%%%%%%%%%%%%%%%%%%%%%%%  
\section{Lepton Sector}
\label{sec:leptons}
 %%%%%%%%%%%%%%%%%%%%%%%%%%%%%%%%%%%%%  

 In the first step of symmetry breaking we preserve a Klein group contained in $\Delta (384)$ together with a CP symmetry, represented by a CP transformation in Eq.~(\ref{eq:CP3}), 
 in the neutrino sector, while a $Z_3$ symmetry is conserved among charged leptons, 
 \begin{equation}
\label{eq:GnuGe} 
 G_{\nu, 1}= Z_2 \times Z_2 \times CP \;\; \mbox{and} \;\; G_l=Z_3 \, ,
\end{equation} 
see figure~\ref{fig1}.
We choose as generators of the Klein group $c^4$ and $a \, b \, c^k$ and leave $k$ as well as the parameter $s$, labelling the CP symmetry, unspecified for the moment. 
 As can be checked explicitly, the generators of the 
Klein group and the CP transformation commute in the representation ${\bf 3}$, i.e. 
\begin{equation}
X ({\bf 3}) (s) \, (c ({\bf 3})^4)^\star - c ({\bf 3})^4 \, X ({\bf 3}) (s) =0 \;\; \mbox{and} \;\; X ({\bf 3}) (s) \, (a ({\bf 3}) \, b ({\bf 3}) \, c ({\bf 3})^k)^\star - a ({\bf 3}) \, b ({\bf 3}) \, c ({\bf 3})^k \, X ({\bf 3}) (s) =0
\end{equation}
with $^\star$ denoting complex conjugation. 
This implies that also the other elements of the Klein group commute with the CP transformation.
For $G_l$ we use the generator $a$. The contributions $U_\nu$ and $U_l$ to the lepton mixing matrix $U_{\mbox{\tiny PMNS}}$, arising from the neutrino and charged lepton sector, respectively, are determined by the unitary matrices diagonalizing  
the generators of the residual symmetries $G_{\nu, 1}$ and $G_l$, respectively.\footnote{We note that we constrain $U_\nu$ additionally such that the CP transformation $X ({\bf 3}) (s)$ becomes the identity 
matrix after applying $U_\nu$, i.e. $U_\nu^\dagger \, X ({\bf 3}) (s) \, U_\nu^\star$ is the identity matrix.} 
They are of the form
\begin{equation}
\label{eq:UnuUlstep1}
U_\nu (k, s) = \frac{1}{\sqrt{2}} \, \left(
\begin{array}{ccc}
\omega_{16}^{s-k} & 0 & -i \, \omega_{16}^{s-k} \\
\omega_{16}^{k+s}  & 0 & i \, \omega_{16}^{k+s} \\
0 & \sqrt{2} \, \omega_8^{8- s} & 0
\end{array}
\right)
\;\; \mbox{and} \;\;
U_l=\frac{1}{\sqrt{3}} \, \left(
\begin{array}{ccc}
1 & \omega & \omega^2\\
1 & \omega^2 & \omega\\
1 & 1 & 1
\end{array}
\right)
\end{equation}
with $\omega_{16}=e^{2 \pi i/16}$ and $\omega=e^{2 \pi i/3}$, 
where we have chosen a particular ordering of the columns, and thus ordering of the neutrino and charged lepton masses, that can entail a viable description of the data on lepton mixing parameters.\footnote{As
is well-known, exchanging the second and third column of $U_l$ leads to the atmospheric mixing angle changing octant and the Dirac phase sign.} 
The form of the lepton mixing matrix is then
\begin{equation}
\label{eq:UPMNSstep1}
U_{\mbox{\tiny PMNS}, 1} (k, s)= U_l^\dagger U_\nu (k, s)= \frac{1}{\sqrt{6}} \, \left(
\begin{array}{ccc}
2 \, \omega_{16}^s \, \cos \phi_k & \sqrt{2} \, \omega_8^{8-s} &  -2 \,  \omega_{16}^s \, \sin  \phi_k\\
 - \omega_{16}^s \, (\cos \phi_k + \sqrt{3} \, \sin \phi_k) & \sqrt{2} \, \omega_8^{8-s} &  -\omega_{16}^s \, (\sqrt{3} \, \cos  \phi_k - \sin  \phi_k) \\
-  \omega_{16}^s \, (\cos \phi_k - \sqrt{3} \, \sin  \phi_k) & \sqrt{2} \, \omega_8^{8-s} &  \omega_{16}^s \, (\sqrt{3} \, \cos  \phi_k + \sin  \phi_k) 
\end{array}
\right) \;\; \mbox{with} \;\; \phi_k=\frac{\pi \, k}{8} \, .
\end{equation} 
We see, in particular, that the sine of the reactor mixing angle $\theta_{13}$ is given by 
\begin{equation}
\sin^2 \theta_{13}= \frac 23 \, \sin^2 \phi_k \, . 
\end{equation}
Taking into account the admitted choices of $k$,
$k=0,1,2,3$, we find that the matrix in Eq.~(\ref{eq:UPMNSstep1}) becomes the TB mixing matrix for the choice\footnote{This result can be understood by considering the group arising
from the combination of the generators of the Klein group and of the $Z_3$ symmetry, preserved among neutrinos and charged leptons, respectively, as one 
generates with the elements $c^4$, $a \, b$ and $a$ the group $\Delta (24)$ ($n=2$) that is isomorphic to $S_4$. The latter group is well-known for leading to TB mixing~\cite{Lam:2008rs}, if a
residual Klein group among neutrinos and a $Z_3$ symmetry in the charged lepton sector is present.} 
\begin{equation}
\label{eq:kchoice}
k=0 \, .
\end{equation}
This can be considered as a reasonable result after the first step of symmetry breaking.
 
  In the second step the Klein group in the neutrino sector is broken to a $Z_2$ symmetry, generated by $a \, b \, c^4$, while the CP symmetry is preserved
  \begin{equation}
  G_{\nu,2}=Z_2 \times CP \, .
  \end{equation}
   The residual symmetry among charged
 leptons remains untouched, see figure~\ref{fig1}. In this way, the contribution to lepton mixing, arising from the neutrino sector, becomes less constrained, i.e. an additional rotation in the $(23)$-plane
 through an angle $\theta$, undetermined by the symmetries of the theory and in the range between $0$ and $\pi$, has to be taken into account. The plane of the rotation is fixed by the fact
 that the unbroken generator $a \, b \, c^4$ in the representation ${\bf 3}$ reads after applying $U_\nu  (k=0, s)$
 \begin{equation}
 \label{eq:formabc4Unu}
 U_\nu (k=0, s)^\dagger \, a ({\bf 3}) \, b ({\bf 3}) \, c ({\bf 3})^4 \, U_\nu (k=0, s) = \mbox{diag} \, (-1,1,1) \, .
 \end{equation}
  We thus arrive at a lepton mixing matrix of the form
 \begin{equation}
 \label{eq:PMNSstep2}
 U_{\mbox{\tiny PMNS}, 2} (s)= U_{\mbox{\tiny PMNS}, 1} (k=0, s) \, R_{23} (\theta) 
 \;\;\; \mbox{with} \;\;\; 
 R_{23} (\theta) = \left(
 \begin{array}{ccc}
 1 & 0 & 0\\
 0 & \cos\theta & \sin\theta\\
 0 & -\sin\theta & \cos\theta
  \end{array}
 \right) \, .
 \end{equation}
  When using $ U_{\mbox{\tiny PMNS}, 2} (s)$ as lepton mixing matrix, we assume that the diagonal elements of the neutrino mass matrix are 
    positive semi-definite after the application of $U_\nu (k=0, s) \, R_{23} (\theta)$. 
    If this is not given, additional signs will appear in the Majorana invariants $I_1$ and $I_2$, defined in Eq.~(\ref{eq:CPinvleptonshere}), compare discussion in~\cite{Feruglio:2012cw,Hagedorn:2014wha}.
 From $ U_{\mbox{\tiny PMNS}, 2} (s)$ we read off that 
 \begin{equation}
 \label{eq:theta13_theta}
\sin^2 \theta_{13} = \frac 13 \, \sin^2 \theta \, .
\end{equation}
 Thus, the experimental data on the reactor mixing angle are accommodated well, $\sin^2 \theta_{13} \approx 0.022$~\cite{Esteban:2016qun}, for\footnote{We note
that there exists a second solution with $\theta$ being replaced by $\pi-\theta$. This is known from preceding analyses~\cite{Hagedorn:2014wha}.}
\begin{equation}
\label{eq:thetavalue}
\theta \approx 0.26 \, .
\end{equation}
Also the value of the solar mixing angle is 
fixed, since it is strongly correlated with the reactor one
\begin{equation}
\label{eq:theta13theta12}
\sin^2 \theta_{12} = \frac{\cos^2 \theta}{2+\cos^2 \theta} = \frac{1}{3} \, \left( \frac{1 -3 \, \sin^2 \theta_{13}}{1 -\sin^2 \theta_{13}} \right) \approx 0.318 \, .
\end{equation}
This sum rule is well-known~\cite{Albright:2008rp} and follows from the fact that the lepton mixing matrix in Eq.~(\ref{eq:PMNSstep2}) leads to $\mbox{TM}_1$ mixing~\cite{Lam:2006wm}.
 In contrast, the atmospheric mixing angle $\theta_{23}$ also depends on the value of $s$
\begin{equation}
\label{eq:sinsqtheta23}
\sin^2 \theta_{23} = \frac{1}{2} \, \left( 1- \left( \frac{2 \, \sqrt{6} \, \sin 2 \, \theta}{5+\cos 2 \, \theta} \right)\, \cos \left( \frac{3 \, \pi \, s}{8} \right) \right) \, .
\end{equation}
 Using Eq.~(\ref{eq:theta13_theta}) and that $\theta$ is small, see Eq.~(\ref{eq:thetavalue}), we can relate $\theta_{23}$ and $\theta_{13}$
\begin{equation}
\label{eq:theta23theta13}
\sin^2 \theta_{23} \approx \frac 12 - \sqrt{2} \, \cos \left(\frac{3 \, \pi \, s}{8} \right) \, \sin\theta_{13} \, .
\end{equation}
This relation can be used to express $\cos \left(\frac{3 \, \pi \, s}{8} \right)$ in terms of the two lepton mixing angles $\theta_{13}$ and $\theta_{23}$. 
Requesting $\sin^2\theta_{23}$ to be within the experimentally preferred $3 \, \sigma$ range~\cite{Esteban:2016qun} for $\theta\approx 0.26$ excludes several choices of $s$: $s=0$, $s=2$, 
$s=3$, $s=5$ and $s=6$, while $s=1$ leads to $\sin^2\theta_{23}$ very close to the lower $3 \, \sigma$ bound. The remaining two values of $s$, $s=4$ and $s=7$, lead to
maximal atmospheric mixing and $\sin^2 \theta_{23} \approx 0.579$, respectively. 

The results for the amount of CP violation in the lepton sector can be quantified
with the Jarlskog invariant $J_{\mbox{\tiny CP}}$ and the Majorana invariants $I_1$ and $I_2$~\cite{Jenkins:2007ip} (see also \cite{Branco:1986gr,Nieves:1987pp,delAguila:1995bk}). These are defined as
\begin{eqnarray}
\nonumber
J_{\mbox{\tiny CP}} &=&{\rm Im} \left[  (U_{\mbox{\tiny PMNS}})_{11} \, (U_{\mbox{\tiny PMNS}})_{13}^\star \, (U_{\mbox{\tiny PMNS}})_{31}^\star \, (U_{\mbox{\tiny PMNS}})_{33} \right] 
= \frac 18 \, \sin 2 \, \theta_{12} \sin 2 \, \theta_{23} \, \sin 2 \, \theta_{13} \, \cos \theta_{13} \, \sin \delta \, ,
\\ \nonumber
I_1 &=& {\rm Im} [(U_{\mbox{\tiny PMNS}})_{12}^2 \, ((U_{\mbox{\tiny PMNS}})_{11}^\star)^2] = \sin^2 \theta_{12} \, \cos^2 \theta _{12} \, \cos^4 \theta_{13} \, \sin \alpha \, ,
\\
\label{eq:CPinvleptonsgen}
I_2 &=&  {\rm Im} [(U_{\mbox{\tiny PMNS}})_{13}^2 \, ((U_{\mbox{\tiny PMNS}})_{11}^\star)^2] = \sin^2 \theta_{13} \, \cos^2 \theta_{12} \, \cos^2 \theta_{13} \, \sin \beta 
\end{eqnarray}
 and read for $U_{\mbox{\tiny PMNS}}=U_{\mbox{\tiny PMNS}, 2} (s)$ in Eq.~(\ref{eq:PMNSstep2})
\begin{equation}
\label{eq:CPinvleptonshere}
J_{\mbox{\tiny CP}} = -\frac{1}{6 \, \sqrt{6}} \, \sin 2 \, \theta \, \sin \left( \frac{3 \, \pi \, s}{8} \right) \; , \;
I_1 =  -\frac 29 \, \cos^2 \theta \, \sin \left( \frac{3 \, \pi \, s}{4} \right) \;\; \mbox{and} \;\;
I_2 = -\frac 29 \, \sin^2 \theta \, \sin \left( \frac{3 \, \pi \, s}{4} \right) \, .
\end{equation}
For $\theta \ll 1$ we can extract simple formulae for the CP phases $\delta$, $\alpha$ and $\beta$
\begin{equation}
\label{eq:CPphasessimple}
\sin \delta \approx - \sin \left( \frac{3 \, \pi \, s}{8} \right) \; , \;\; \sin\alpha = \sin\beta =  - \sin \left( \frac{3 \, \pi \, s}{4} \right) \, ,
\end{equation}
compare also~\cite{Hagedorn:2014wha}.
 With the help of Eq.~(\ref{eq:theta23theta13}) and the first relation in Eq.~(\ref{eq:CPphasessimple}) we can relate the Dirac phase $\delta$ and the two lepton mixing angles $\theta_{13}$
and $\theta_{23}$
\begin{equation}
\label{eq:delta_theta23theta13} 
\left( 1- 2 \, \sin^2 \theta_{23} \right)^2 \approx 8 \, \sin^2 \theta_{13} \, \cos^2 \delta \, .
\end{equation}
This sum rule turns out to be a very good approximation of the exact one, given in~\cite{Albright:2008rp} for $\mbox{TM}_1$ mixing. 
Furthermore, we can express the Majorana phases $\alpha$ and $\beta$ in terms of the Dirac phase $\delta$ and the two lepton mixing angles $\theta_{13}$ and $\theta_{23}$
\begin{equation}
\label{eq:alphabeta_deltatheta13theta23}
\sin\alpha=\sin\beta= -2 \, \sin \left( \frac{3 \, \pi \, s}{8} \right) \, \cos \left( \frac{3 \, \pi \, s}{8} \right) \approx \sin\delta \, \left( \frac{1-2 \, \sin^2 \theta_{23}}{\sqrt{2} \, \sin\theta_{13}} \right)  \, .
\end{equation}
Using Eq.~(\ref{eq:CPinvleptonshere}) and $\theta \approx 0.26$ together with the choice 
 $s=7$, we find
\begin{equation}
\label{eq:CPphasesleptons7}
\sin \delta  \approx -0.936 \;\; , \; \sin\alpha = \sin\beta = \frac{1}{\sqrt{2}}  \; .
\end{equation}
The value of $\delta$ is hence close to the best fit value, obtained from the global fit in~\cite{Esteban:2016qun}. The choice $s=4$ would in contrast correspond to maximal CP violation $\delta=\pi/2$
  which is disfavoured at more than the $3 \, \sigma$ level.\footnote{A possibility to cure this is to include an additional permutation of the second and
 third generation of charged leptons. Since the atmospheric mixing angle is maximal for $s=4$, this permutation does not affect its value.} We thus consider 
 \begin{equation}
 s=7 
  \end{equation}
  as most suitable choice for $s$. Using the information on the two Majorana phases $\alpha$ and $\beta$, we can compute
the range of the quantity $m_{ee}$, accessible in neutrinoless double beta decay. 
  Given that this scenario does not make predictions
for the neutrino mass spectrum, we parametrize the latter in terms of the lightest neutrino mass $m_0$, $0 \leq m_0 \lesssim 4 \times 10^{-2} \, \mbox{eV}$ in agreement with limits from cosmology~\cite{Aghanim:2018eyx}, and the best fit values of the two mass squared differences $\Delta m_{\mbox{\footnotesize sol}}^2$ and $\Delta m_{\mbox{\footnotesize atm}}^2$, which we take for the two different neutrino mass orderings from~\cite{Esteban:2016qun}.
Assuming neutrinos follow normal ordering, we obtain
\begin{equation}
\label{eq:mees7}
2.86 \times 10^{-3} \, \mbox{eV} \lesssim m_{ee} \lesssim 1.94 \times 10^{-2} \, \mbox{eV} \;\;\; \mbox{and} \;\;\; 2.43 \times 10^{-2} \, \mbox{eV} \lesssim m_{ee} \lesssim 3.11 \times 10^{-2} \, \mbox{eV}
\end{equation}
for neutrino masses with inverted ordering. We note, in particular, that for neutrinos with normal ordering $m_{ee}$ cannot vanish and has a minimum value for $m_0 \approx 3.62 \times 10^{-3} \, \mbox{eV}$ and that for inverted ordering the minimum of $m_{ee}$ is also larger than the lower bound expected when using the current information from neutrino oscillation experiments at the $3 \, \sigma$ level, cf.~also figure 9  in~\cite{Hagedorn:2016lva}.

 This mixing pattern belongs, according to the classification in~\cite{Hagedorn:2014wha}, to Case 3 b.1). In~\cite{Hagedorn:2014wha} very similar numerical results have been found for $\Delta (6 \, n^2)$ with $n=8$, 
 $G_l=Z_3$ generated by $a$ and $Z_2 \times CP$, preserved in the neutrino sector, with $b \, c^4 \, d^4$ being the generator of $Z_2$ and the CP transformation given as 
   $b ({\bf 3}) \, c ({\bf 3})^7 \, d ({\bf 3}) \, X_0 ({\bf 3})$ in the representation ${\bf 3}$. This choice of residual symmetries is related to the one used in the present study by the similarity transformation $a$,
  see~\cite{Hagedorn:2014wha} for details, and thus the results coincide, up to a different definition of the angle $\theta$.
  
  As all data on lepton mixing can be satisfactorily described and predictions for CP phases are made after the second step of symmetry breaking, we conclude the study of the lepton sector at this point.
  We note, however, that in concrete models that realize this symmetry breaking sequence we expect that higher order corrections at a certain point break all residual symmetries, see e.g.~\cite{Feruglio:2013hia,model}.

 %%%%%%%%%%%%%%%%%%%%%%%%%%%%%%%%%%%%%  
\section{Quark Sector}
\label{sec:quarks}
 %%%%%%%%%%%%%%%%%%%%%%%%%%%%%%%%%%%%%  

In the first step of symmetry breaking the same residual symmetry is preserved in the up quark sector as in the neutrino one and a $Z_{16}$ subgroup remains intact in
the down quark sector, i.e.
\begin{equation}
G_{u, 1}=G_{\nu, 1}=Z_2 \times Z_2 \times CP \;\;\; \mbox{and} \;\;\; G^{\pm}_{d, 1}=Z_{16} \, ,
\end{equation}
as also shown in figure~\ref{fig1}.
In order to obtain a value for the Cabibbo angle close to the measured one at this step of symmetry breaking, we have to choose the generator of $G^\pm_{d, 1}$
as $a \, b \, d^{2 \, (4-k) \pm 1}$ 
 with $k=0,1,2,3$ for $G_{u, 1}=G_{\nu, 1}$. The form of the contribution of the up quark sector to the quark mixing matrix
is the same as $U_\nu (k, s)$ in Eq.~(\ref{eq:UnuUlstep1}), up to the ordering of the columns of the mixing matrix that depends on the up quark masses. Indeed, we choose in the following a slightly different
ordering for the columns of $U_u (k, s)$ than of $U_\nu (k, s)$, i.e. 
\begin{equation}
\label{eq:Uustep1}
U_u (k, s)= \frac{1}{\sqrt{2}} \, \left(
\begin{array}{ccc}
\omega_{16}^{s-k} &  -i \, \omega_{16}^{s-k} & 0 \\
\omega_{16}^{k+s}  & i \, \omega_{16}^{k+s} & 0 \\
0 & 0 & \sqrt{2} \, \omega_8^{8- s} 
\end{array}
\right) \, .
\end{equation}
In addition, the form of the mixing matrix $U_d$ that encodes the contribution from the down quark sector to the quark mixing matrix is
\begin{equation}
\label{eq:Udstep1}
U_{d, \pm 1} (k)=\frac{1}{\sqrt{2}} \, \left(
\begin{array}{ccc}
\omega_{16}^{\pm 1-2\, k} & -\omega_{16}^{\pm 1-2\, k}  & 0\\
1 & 1 & 0\\
0 & 0 & \sqrt{2}
\end{array}
\right)
\, ,
\end{equation}
where $U_{d, \pm 1} (k)$ 
 corresponds to the choice of $G_{d, 1}^{\pm}$ 
  as residual symmetry among down quarks. 
The absolute values of the elements of the resulting quark mixing matrix read for both choices 
\begin{equation}
\label{eq:VCKMabsstep1}
\left| V_{\mbox{\tiny CKM}, 1} \right|=\left| U_u (k, s)^\dagger \, U_{d, \pm 1} (k)\right| = \left(
\begin{array}{ccc}
\cos \pi/16 & \sin \pi/16 & 0\\
\sin \pi/16 & \cos \pi/16 & 0\\
0 & 0 & 1
\end{array}
\right) 
\end{equation}
and are independent of $k$.
As can be read off, the size of the Cabibbo angle $\theta_C$ is $\sin \pi/16 \approx 0.195$ which represents a reasonably good leading order  
approximation to the experimentally measured value~\cite{PDG2018}. Combining the generators of $G_{u, 1}$ and $G_{d, 1}^{\pm}$ leads for all choices of $k$ to a group that has 128 elements and is called [[128,67]]
in the computer program GAP~\cite{GAP1,GAP2}. This group can be written as $(Z_8 \times Z_8) \rtimes Z_2$ and only has irreducible representations of dimension one and two. We have explicitly checked that
the representation ${\bf 3}$ of $\Delta (384)$ decomposes into a complex one- and a faithful, irreducible, complex two-dimensional representation in the group [[128,67]]. 
 The observation that breaking $\Delta (384)$ to a $Z_{16}$ subgroup and a Klein group can lead to one mixing angle $\sin\pi/16 \approx 0.195$, suitable for a leading order description of quark mixing,
 has already been made in~\cite{deAdelhartToorop:2011re}.
 
In the second step of symmetry breaking in the quark sector the leading order result of the Cabibbo angle is brought into full agreement with the experimental data~\cite{PDG2018}. This can be
achieved in two different ways. One possibility is that the residual symmetry $G_{u, 1}$ is broken to 
\begin{equation}
\label{eq:Gustep2}
G_{u, 2}=Z_2 \times CP \, ,
\end{equation}
where the remaining $Z_2$ symmetry is generated by $c^4$, see figure~\ref{fig1}. The latter element reads in the representation ${\bf 3}$ after applying the matrix $U_u (k, s)$
\begin{equation}
\label{eq:c4Uu}
U_u (k, s)^\dagger \, c ({\bf 3})^4 \, U_u (k, s)= \mbox{diag} \, (-1,-1,1) \, ,
\end{equation}
meaning that it allows for a rotation $R_{12} (\theta_u)$ in the (12)-plane through an angle $\theta_u$.\footnote{Like in the case of $G_{\nu, 2}$, the fact that $G_{u, 2}$ contains CP
as symmetry constrains the additional unitary matrix, leaving Eq.~(\ref{eq:c4Uu}) invariant, to be a rotation with one free real parameter only.}  
  The absolute values of the elements of the resulting quark mixing matrix become
\begin{equation}
\label{eq:VCKMabsstep2u}
|V_{\mbox{\tiny CKM}, 2, u}| = |(U_u (k, s)\, R_{12} (\theta_u))^\dagger U_{d, \pm 1} (k)| = \left(
\begin{array}{ccc}
|\cos (\pi/16 \mp \theta_u)| & |\sin (\pi/16 \mp \theta_u)| & 0\\
|\sin  (\pi/16 \mp \theta_u)|  & |\cos  (\pi/16 \mp \theta_u)|  & 0\\
0 & 0 & 1
\end{array}
\right) \, .
\end{equation}
Choosing $\theta_u \approx \mp 0.030$ depending on the sign in Eq.~(\ref{eq:VCKMabsstep2u}), 
we achieve $|V_{us}| = 0.22452$, corresponding to the experimental best fit value~\cite{PDG2018}.
The other possibility is to reduce $G_{d, 1}^{\pm}$ to\footnote{We could also consider $G_{d, 2}^\pm=Z_4$ or $G_{d, 2}^\pm=Z_2$, since these choices will lead to the same additional matrix $U_{12} (\theta_d, \psi_d)$,
 as also their generators have two degenerate eigenvalues in the representation ${\bf 3}$.} 
\begin{equation}
G_{d, 2}^{\pm}= Z_8 \, ,
\end{equation}
generated by $(a \, b \, d^{2 \, (4-k) \pm 1})^2=(c \, d^2)^{\pm 1-2 \, k}$, 
 respectively, while the residual symmetry $G_{u, 1}$ in the up quark sector remains intact. The form of  
these generators in the representation ${\bf 3}$ is after applying $U_{d, \pm 1} (k)$ 
\begin{equation}
U_{d, \pm 1} (k)^\dagger \, (c ({\bf 3}) \, d ({\bf 3})^2)^{\pm 1-2 \, k} \, U_{d, \pm 1} (k)= 
\left( \begin{array}{ccc}
 \omega_8^{\pm 1 -2 \, k} & 0 & 0\\
0 &  \omega_8^{\pm 1 -2 \, k} & 0\\
0 & 0 & i \, (-1)^{k+ (1\pm 1)/2}
\end{array}
\right) \, .
\end{equation}
This shows that two of their eigenvalues are degenerate for all admitted values of $k$, such that the contribution of the down quark sector to quark mixing includes 
an additional unitary transformation $U_{12} (\theta_d, \psi_d)$ of the form
\begin{equation}
\label{eq:U12dform}
U_{12} (\theta_d, \psi_d)= \left(
\begin{array}{ccc}
\cos \theta_d & \sin \theta_d \, e^{- i \, \psi_d} & 0 \\
- \sin \theta_d \, e^{i \, \psi_d} & \cos \theta_d & 0 \\
0 & 0 & 1
\end{array}
\right)
\end{equation}
with free parameters $\theta_d$ and $\psi_d$.
 The resulting quark mixing matrix is 
\begin{eqnarray}
\nonumber
&&\left|V_{\mbox{\tiny CKM}, 2, d}\right| =\left| U_u (k, s)^\dagger U_{d, \pm 1} (k) \, U_{12} (\theta_d, \psi_d)\right| = \left(
\begin{array}{ccc}
|\cos \zeta| & |\sin \zeta| & 0\\
|\sin \zeta|  & |\cos \zeta|  & 0\\
0 & 0 & 1
\end{array}
\right) 
\\ \label{eq:VCKMabsstep2d}
\mbox{with} \;\;&&\cos 2 \, \zeta = \cos \pi/8 \, \cos 2 \, \theta_d \mp \sin \pi/8 \, \sin 2 \,\theta_d \, \sin \psi_d \, .
\end{eqnarray}
  This shows that effectively only one real parameter is responsible for achieving $|V_{us}|$ in full accordance with the
experimental results and possible numerical values of $\theta_d$ and $\psi_d$ can be obtained in a similar way as for $\theta_u$. 
  We thus conclude that both versions of the second step of symmetry breaking in the quark sector, together with the two different possible choices $G_{d, 1}^{-}$ and $G^{+}_{d, 1}$, lead to the same phenomenology.
For reasons of minimality, we do not consider the case where $G_{u, 1}$ and $G_{d, 1}^{\pm}$ are both broken to $G_{u, 2}$ and $G_{d, 2}^{\pm}$, respectively. 

In the third step of symmetry breaking the two remaining quark mixing angles are generated together with the CP phase such that all experimental data can be accommodated well~\cite{PDG2018}.
We thus assume that this step of symmetry breaking induces in the quark mixing matrix two additional angles acting in the $(13)$- and $(23)$-planes. 
We can envisage different scenarios for this last step:
$a)$ only the residual flavor symmetry in the up quark sector is broken, while leaving the CP
symmetry in this sector together with the residual group among down quarks intact,
$b)$ the residual symmetry in the up quark sector is broken completely, while the one in the down quark sector is still preserved, $c)$ the symmetry in the up quark sector remains
untouched, while the residual one among down quarks is broken, as well as $d)$ the residual symmetries in both, up and down quark, sectors are broken.

With one of the four different quark mixing matrices, that can possibly arise from the second step of symmetry breaking, see Eqs.~(\ref{eq:VCKMabsstep2u}) and (\ref{eq:VCKMabsstep2d}), 
the different versions of the third step lead to one of the 
following results for quark mixing: if the residual symmetry is only broken in one of the two sectors in the third step of symmetry breaking, we find that $J_{\mbox{\tiny CP}}^q$ crucially depends on the
phases that are contained in the unitary matrices, arising from the last step of symmetry breaking (and possibly the second one, if in that the residual symmetry of the down quark sector
is reduced). The value of such phases is in general not constrained by the residual symmetries. However, in a concrete model these can take specific values due to fixed phases, originating
from the vacuum expectation values of some flavor (and CP) symmetry breaking fields, see~\cite{model}.
 One example is to assume that in the second and third step only the residual symmetry in the up quark sector is broken, i.e.~after the second step the quark mixing matrix is
of the form as in Eq.~(\ref{eq:VCKMabsstep2u}) and scenario $b)$ is realized in the third step. We find
\begin{eqnarray}
\nonumber
&&\sin\theta_{12}^q \approx \left| \sin \left( \frac{\pi}{16} \mp \theta_u \right)\right| \; , \;\; \theta_{23}^q \approx \left| \theta_{u, 23} \right| \; , \;\; \theta_{13}^q \approx \left| \theta_{u, 13} \right| \; ,
\\ \label{eq:analyticsquarksstep3noks}
\mbox{and} \;\;&&J_{\mbox{\tiny CP}}^q = \pm \frac{1}{8} \, \sin \left( \frac{\pi}{8} \mp 2 \theta_u \right) \, \sin 2 \theta_{u, 13} \, \sin 2 \theta_{u, 23} \, \cos \theta_{u, 23} \, \sin \psi
\;\; \mbox{with} \;\; \psi = \psi_{u, 13} - \psi_{u, 23} \, , 
\end{eqnarray}
where $\theta_{u, 13}$, $\psi_{u, 13}$ and $\theta_{u, 23}$, $\psi_{u, 23}$ parametrize the unitary transformations, arising from the third step of symmetry breaking, analogously to $\theta_d$, $\psi_d$ for 
the matrix $U_{12} (\theta_d, \psi_d)$ in Eq.~(\ref{eq:U12dform}). We note that only the difference $\psi$ of the two phases $\psi_{u, 13}$ and $\psi_{u, 23}$ enters and that its value is essential for the determination
of the amount of CP violation in the quark sector.
In order to accommodate the best fit values for the quark mixing angles $\theta_{ij}^q$ and the Jarlskog invariant \cite{PDG2018}, we choose the parameters as\footnote{The signs of $\theta_{u, 23}$ and $\theta_{u, 13}$ are not uniquely fixed, but constrained through the requirement to eventually obtain the correct sign for $J_{\mbox{\tiny CP}}^q$ for a fixed value of $\psi$.}
\begin{equation}  
\label{eq:parametersstep3noks}
\theta_u \approx \mp 0.030 \; , \;\; \left|\theta_{u, 23} \right| \approx 0.042 \; , \;\; \left| \theta_{u, 13}\right| \approx 0.00365 \;\; \mbox{and} \;\; \left|\sin\psi\right| \approx 0.95 \, ,
\end{equation}
showing that the breaking in the second and third step is small. Indeed, it is almost an order of magnitude smaller than in the lepton sector, compare Eq.~(\ref{eq:thetavalue}).

If instead one of the remaining quark mixing angles comes from the breaking of the residual symmetry in the up quark and the other one from that in the down quark sector,
$J_{\mbox{\tiny CP}}^q$ always depends on the choice of the parameters $k$ and $s$. 
Minimal viable cases, in which only one phase is due to the symmetry breaking, are encountered, if the second step of symmetry breaking takes place in the up quark sector. Out of the four possible cases, depending on the choice of $G_{d, 1}^{\pm}$ and on whether $\theta_{23}^q$ or $\theta_{13}^q$ is dominantly generated in the up quark sector, two are particularly interesting. In these cases $\theta_{23}^q$ is due to the symmetry breaking in the up quark sector. They lead to a viable quark mixing matrix, even if no phase is induced by the  symmetry breaking in the down quark sector in the third step. We display this symmetry breaking scenario 
in figure~\ref{fig1}. We get as result for the mixing angles and $J_{\mbox{\tiny CP}}^q$ 
\begin{eqnarray}
\nonumber
&&\sin\theta_{12}^q \approx \left| \sin  \left( \frac{\pi}{16} \mp \theta_u \right)\right| \; , \;\; \theta_{23}^q \approx \left| \theta_{u, 23} \right| \; , \;\; \theta_{13}^q \approx \left| \theta_{d, 13} \, \cos \left( \frac{\pi}{16} \mp \theta_u \right) \right| \; ,
\\ \label{eq:analyticsquarksstep3ks}
&&J_{\mbox{\tiny CP}}^q = \mp \, \frac{1}{8} \, \sin \left( \frac{\pi}{8} \mp 2 \theta_u \right) \, \cos \left( \frac{\pi}{16} \mp \theta_u \right) \, \sin 2 \theta_{d, 13} \, \sin 2 \theta_{u, 23} \, \sin \left(  \phi_k + \frac{3 \, \pi \, s}{8} \mp \frac{\pi}{16} + \psi_{d, 13} \right) 
\end{eqnarray}
 depending on $G_{d, 1}^\pm$ .
Extracting a formula for the CP phase $\delta^q$ itself we find
\begin{equation}
\label{eq:formulasindq}
|\sin\delta^q| \approx \left|\sin \left( \phi_k + \frac{3 \, \pi \, s}{8} \mp \frac{\pi}{16} + \psi_{d, 13} \right)\right| \, .
\end{equation}
 This relation is very interesting, since it allows to relate the CP phase $\delta^q$ with the lepton mixing parameters $\delta$, $\theta_{13}$ and $\theta_{23}$. 
Since $k$ is set to zero after the first step of symmetry breaking in the lepton sector, see Eq.~(\ref{eq:kchoice}), we focus on this case. Using Eq.~(\ref{eq:theta23theta13}) 
and the first relation in Eq.~(\ref{eq:CPphasessimple}), we find
\begin{eqnarray}\nonumber
|\sin\delta^q| &\approx& \left|  \sin \left(  \frac{3 \, \pi \, s}{8} \right) \, \cos \left( \mp \frac{\pi}{16} + \psi_{d,13}  \right) + \cos \left(  \frac{3 \, \pi \, s}{8} \right) \, \sin \left( \mp \frac{\pi}{16} + \psi_{d,13}  \right) \right|
\\ \label{eq:deltaq_relk0gen}
&\approx&  \left|  \sin\delta  \, \cos \left( \theta_C \mp \psi_{d,13}  \right) \pm \left(\frac{1-2 \, \sin^2 \theta_{23}}{2 \, \sqrt{2} \, \sin \theta_{13}}  \right)\, \sin \left( \theta_C \mp \psi_{d,13}  \right)  \right| \, ,
\end{eqnarray}
where we have also inserted the Cabibbo angle $\theta_C$, whose value is given by $\theta_C =\sin\pi/16 \approx \pi/16$ after the first step of symmetry breaking in the quark sector, compare Eq.~(\ref{eq:VCKMabsstep1}).
This relation simplifies further, if we consider a case, where $\psi_{d,13}$ vanishes, namely
\begin{equation}
\label{eq:deltaq_relk0psid130}
|\sin\delta^q| \approx \left| \sin\delta  \, \cos \left( \theta_C \right) \pm \left(\frac{1-2 \, \sin^2 \theta_{23}}{2 \, \sqrt{2} \, \sin \theta_{13}}  \right)\, \sin \left(\theta_C \right)  \right| \, .
\end{equation}  
The experimental best fit values of the quark mixing angles are achieved for 
\begin{equation}
\label{eq:valuesangles}
\theta_u \approx \mp 0.030 \; , \;\; \left|\theta_{u, 23}\right| \approx 0.042 \;\; \mbox{and} \;\; \left|\theta_{d, 13}\right| \approx 0.00375 \, ,
\end{equation}
demonstrating that also in this case only a small breaking is needed at the second and third step in the quark sector.  
Using the values of $k$ and $s$ that are most suitable for the description of the lepton sector, $k=0$ and $s=7$, and choosing the signs of the parameters $\theta_{u, 23}$ and $\theta_{d, 13}$ accordingly, we arrive at
\begin{equation}
\label{eq:valuesJCPqwithpsi}
J_{\mbox{\tiny CP}}^q \approx 3.35 \times 10^{-5} \times \, \cos \left( \left( 2 \mp 1\right) \, \frac{\pi}{16} + \psi_{d, 13} \right) \, .
\end{equation}
This formula shows that even in the limit, in which no phase $\psi_{d, 13}$ is generated in the third step of symmetry breaking, the value of $J_{\mbox{\tiny CP}}^q$
is correctly accommodated
\begin{equation}
\label{eq:valuesJCPqnopsi}
J_{\mbox{\tiny CP}}^q \approx 3.29 \times 10^{-5} \;\; \mbox{for} \;\; G^+_{d,1} \;\; \mbox{and} \;\; J_{\mbox{\tiny CP}}^q \approx 2.79 \times 10^{-5} \;\; \mbox{for} \;\; G^-_{d,1}
\end{equation}
that is in the experimentally preferred $1 \, \sigma$ range for $G^+_{d,1}$ and $3 \, \sigma$ range for $G^-_{d,1}$, compare $J_{\mbox{\tiny CP}}^q= \left( 3.18 \pm 0.15 \right) \times 10^{-5}$~\cite{PDG2018}. 
The vanishing of the phase $\psi_{d, 13}$ can be explained in a concrete model, if e.g.~the operators relevant for the symmetry breaking in the down quark sector do only contain
flavor (and CP) symmetry breaking fields that acquire real vacuum expectation values. In order to achieve the current experimental best fit value of $J_{\mbox{\tiny CP}}^q$, 
  the phase $\psi_{d, 13}$ should take a value
\begin{equation} 
\psi_{d, 13} \approx 0.13 \;\; \mbox{for} \;\; G^+_{d,1} \;\; \mbox{and} \;\; \psi_{d, 13} \approx -0.27 \;\; \mbox{for} \;\; G^-_{d,1} \, ,
\end{equation}
respectively. All shown numerical estimates agree well with the results of a $\chi^2$ analysis, based on the experimental best fit values and $1 \, \sigma$ errors found in~\cite{PDG2018}. In this $\chi^2$ analysis we have also
studied quark mixing, arising from the other possible realizations of the second and third symmetry breaking step, and found that all of them can lead to a viable description of the experimental data except for one case, 
where no CP violation can be induced, since the only symmetry that is reduced in the second and third step is the residual flavor symmetry in the up quark sector.

We have thus shown that quark mixing can be successfully described with three steps of symmetry breaking. While the breaking in the first step is unique, different options arise at the second and third steps.
As explained, these can lead to different results. In particular, the case, where only the residual symmetry in either the up quark or the down quark sector is broken at the third step, has to be distinguished from the case, where 
symmetry breaking occurs in both sectors in the third step, see figure~\ref{fig1}. In the former case $J_{\mbox{\tiny CP}}^q$ depends on phases that originate from the (second and) third step of symmetry breaking, whereas in the latter $J_{\mbox{\tiny CP}}^q$ crucially depends on the parameters $k$ and $s$ that determine the amount of CP violation in the lepton sector. This second case thus leads to a sum rule between the CP phase $\delta^q$ and the Cabibbo angle $\theta_C$ in the quark sector and the Dirac phase $\delta$ and the mixing angles $\theta_{13}$ and $\theta_{23}$
in the lepton sector, see Eqs.~(\ref{eq:deltaq_relk0gen}) and (\ref{eq:deltaq_relk0psid130}). In the limit, in which the second and third step of symmetry breaking do not introduce any phases as free parameters, the most suitable choice of $k$ and $s$ for leptons can indeed lead to $J_{\mbox{\tiny CP}}^q$ in accordance with experimental data.  
 Corrections to the entertained symmetry breaking scenario might occur in an explicit model e.g.~through higher-dimensional operators, but are usually suppressed.

%%%%%%%%%%%%%%%%%%%%%%%%%%%%%%%%%%%%%
\section{Summary and Conclusions}
\label{sec:conclusions}
%%%%%%%%%%%%%%%%%%%%%%%%%%%%%%%%%%%%%   

 We have presented a scenario with a flavor and a CP symmetry, where the mixing patterns for  
leptons and quarks arise from the stepwise breaking of these symmetries to different residual subgroups in the different sectors of the theory.
 In particular, the lepton mixing pattern originates from two steps of symmetry breaking, where the first one gives rise 
 to TB mixing and the second one introduces one real quantity, whose size is determined by the requirement to correctly
  accommodate the reactor mixing angle. 
   Furthermore, a sum rule is derived that relates the Dirac and Majorana phases with the reactor and atmospheric mixing angles. 
  One CP symmetry is singled out by the experimental data on the atmospheric mixing angle and the
  indications for $\delta$ close to $3 \, \pi/2$. In the quark sector the Cabbibo angle is fixed to $\theta_C = \sin \pi/16 \approx 0.195$
   after the first step of symmetry breaking and brought into full accordance with experimental data with the second one. In the third step eventually 
  the two other quark mixing angles are generated. For certain sequences of symmetry breaking we find a close correlation between
  the amount of CP violation in the lepton and in the quark sector. This correlation can be expressed as a sum rule among the CP phase $\delta^q$ and the Cabibbo angle $\theta_C$ in the quark sector and the Dirac phase $\delta$ and 
  the reactor and atmospheric mixing angles $\theta_{13}$ and $\theta_{23}$ in the lepton sector.
   The most constrained viable sequence is shown in figure~\ref{fig1},
  where determining the generators of the residual symmetries, common to the neutrino and the up quark sector, with the help of neutrino
  oscillation data can lead to a value of $J_{\mbox{\tiny CP}}^q$ that lies in the experimentally preferred $1 \, \sigma$ range, independent
  of further free parameters.

  A realization of this symmetry-based scenario in a concrete model will be presented elsewhere~\cite{model}. This model will have salient features
  beyond the successful description of lepton and quark mixing in terms of residual symmetries and their breaking, such as the explanation of the charged fermion mass hierarchies via higher-dimensional 
   operators (fixing the ordering of rows and columns of the mixing matrices), the generation of light neutrino masses via the type-I
   seesaw mechanism as well as the spontaneous breaking of the flavor and CP symmetry.

%%%%%%%%%%%%%%%%%%%%%%%%%%%%%%%%%%%%%%  
\section*{Acknowledgments}
%%%%%%%%%%%%%%%%%%%%%%%%%%%%%%%%%%%%%%  

The CP3-Origins centre is partially funded by the Danish National Research Foundation, grant number DNRF90.

%%%%%%%%%%%%%%%%%%%%%%%%%%%%%%%%%%%%%  

 %%%%%%%%%%%%%%%%%%%%%%%%%%%%%%%%%%%%%  


\begin{thebibliography}{10}

\bibitem{Altarelli:2010gt}
Guido Altarelli and Ferruccio Feruglio.
\newblock {Discrete Flavor Symmetries and Models of Neutrino Mixing}.
\newblock {\em Rev. Mod. Phys.}, 82:2701--2729, 2010, 1002.0211.

\bibitem{Ishimori:2010au}
Hajime Ishimori, Tatsuo Kobayashi, Hiroshi Ohki, Yusuke Shimizu, Hiroshi Okada,
  and Morimitsu Tanimoto.
\newblock {Non-Abelian Discrete Symmetries in Particle Physics}.
\newblock {\em Prog. Theor. Phys. Suppl.}, 183:1--163, 2010, 1003.3552.

\bibitem{King:2013eh}
Stephen~F. King and Christoph Luhn.
\newblock {Neutrino Mass and Mixing with Discrete Symmetry}.
\newblock {\em Rept. Prog. Phys.}, 76:056201, 2013, 1301.1340.

\bibitem{Grimus:2011fk}
Walter Grimus and Patrick~Otto Ludl.
\newblock {Finite flavour groups of fermions}.
\newblock {\em J. Phys.}, A45:233001, 2012, 1110.6376.

\bibitem{Lam:2007qc}
C.~S. Lam.
\newblock {Symmetry of Lepton Mixing}.
\newblock {\em Phys. Lett.}, B656:193--198, 2007, 0708.3665.

\bibitem{Blum:2007jz}
A.~Blum, C.~Hagedorn, and M.~Lindner.
\newblock {Fermion Masses and Mixings from Dihedral Flavor Symmetries with
  Preserved Subgroups}.
\newblock {\em Phys. Rev.}, D77:076004, 2008, 0709.3450.

\bibitem{Lam:2008rs}
C.~S. Lam.
\newblock {Determining Horizontal Symmetry from Neutrino Mixing}.
\newblock {\em Phys. Rev. Lett.}, 101:121602, 2008, 0804.2622.

\bibitem{Harrison:2002kp}
P.~F. Harrison and W.~G. Scott.
\newblock {Symmetries and generalizations of tri - bimaximal neutrino mixing}.
\newblock {\em Phys. Lett.}, B535:163--169, 2002, hep-ph/0203209.

\bibitem{Grimus:2003yn}
Walter Grimus and Luis Lavoura.
\newblock {A Nonstandard CP transformation leading to maximal atmospheric
  neutrino mixing}.
\newblock {\em Phys. Lett.}, B579:113--122, 2004, hep-ph/0305309.

\bibitem{Feruglio:2012cw}
Ferruccio Feruglio, Claudia Hagedorn, and Robert Ziegler.
\newblock {Lepton Mixing Parameters from Discrete and CP Symmetries}.
\newblock {\em JHEP}, 07:027, 2013, 1211.5560.

\bibitem{deAdelhartToorop:2011re}
Reinier de~Adelhart~Toorop, Ferruccio Feruglio, and Claudia Hagedorn.
\newblock {Finite Modular Groups and Lepton Mixing}.
\newblock {\em Nucl. Phys.}, B858:437--467, 2012, 1112.1340.

\bibitem{King:2013vna}
Stephen~F. King, Thomas Neder, and Alexander~J. Stuart.
\newblock {Lepton mixing predictions from $\Delta(6 \, n^2)$ family symmetry}.
\newblock {\em Phys. Lett.}, B726:312--315, 2013, 1305.3200.

\bibitem{Hagedorn:2013nra}
Claudia Hagedorn, Aurora Meroni, and Lorenzo Vitale.
\newblock {Mixing patterns from the groups $\Sigma(n \, \varphi)$}.
\newblock {\em J. Phys.}, A47:055201, 2014, 1307.5308.

\bibitem{Fonseca:2014koa}
Renato~M. Fonseca and Walter Grimus.
\newblock {Classification of lepton mixing matrices from finite residual
  symmetries}.
\newblock {\em JHEP}, 09:033, 2014, 1405.3678.

\bibitem{Hagedorn:2014wha}
Claudia Hagedorn, Aurora Meroni, and Emiliano Molinaro.
\newblock {Lepton mixing from $\Delta (3 \, n^2)$ and $\Delta (6 \, n^2)$ and
  CP}.
\newblock {\em Nucl. Phys.}, B891:499--557, 2015, 1408.7118.

\bibitem{Ding:2014ora}
Gui-Jun Ding, Stephen~F. King, and Thomas Neder.
\newblock {Generalised CP and $\Delta(6 \, n^2)$ family symmetry in semi-direct
  models of leptons}.
\newblock {\em JHEP}, 12:007, 2014, 1409.8005.

\bibitem{Talbert:2014bda}
Jim Talbert.
\newblock {[Re]constructing Finite Flavour Groups: Horizontal Symmetry Scans
  from the Bottom-Up}.
\newblock {\em JHEP}, 12:058, 2014, 1409.7310.

\bibitem{Everett:2015oka}
Lisa~L. Everett, Todd Garon, and Alexander~J. Stuart.
\newblock {A Bottom-Up Approach to Lepton Flavor and CP Symmetries}.
\newblock {\em JHEP}, 04:069, 2015, 1501.04336.

\bibitem{Holthausen:2013vba}
Martin Holthausen and Kher~Sham Lim.
\newblock {Quark and Leptonic Mixing Patterns from the Breakdown of a Common
  Discrete Flavor Symmetry}.
\newblock {\em Phys. Rev.}, D88:033018, 2013, 1306.4356.

\bibitem{Araki:2013rkf}
Takeshi Araki, Hiroyuki Ishida, Hajime Ishimori, Tatsuo Kobayashi, and Atsushi
  Ogasahara.
\newblock {CKM matrix and flavor symmetries}.
\newblock {\em Phys. Rev.}, D88:096002, 2013, 1309.4217.

\bibitem{Yao:2015dwa}
Chang-Yuan Yao and Gui-Jun Ding.
\newblock {Lepton and Quark Mixing Patterns from Finite Flavor Symmetries}.
\newblock {\em Phys. Rev.}, D92(9):096010, 2015, 1505.03798.

\bibitem{Varzielas:2016zuo}
Ivo de~Medeiros~Varzielas, Rasmus~W. Rasmussen, and Jim Talbert.
\newblock {Bottom-Up Discrete Symmetries for Cabibbo Mixing}.
\newblock {\em Int. J. Mod. Phys.}, A32(06n07):1750047, 2017, 1605.03581.

\bibitem{Li:2017abz}
Cai-Chang Li, Jun-Nan Lu, and Gui-Jun Ding.
\newblock {Toward a unified interpretation of quark and lepton mixing from
  flavor and CP symmetries}.
\newblock {\em JHEP}, 02:038, 2018, 1706.04576.

\bibitem{Lu:2018oxc}
Jun-Nan Lu and Gui-Jun Ding.
\newblock {Quark and lepton mixing patterns from a common discrete flavor
  symmetry with a generalized CP symmetry}.
\newblock {\em Phys. Rev.}, D98(5):055011, 2018, 1806.02301.

\bibitem{Grimus:1995zi}
W.~Grimus and M.~N. Rebelo.
\newblock {Automorphisms in gauge theories and the definition of CP and P}.
\newblock {\em Phys. Rept.}, 281:239--308, 1997, hep-ph/9506272.

\bibitem{Holthausen:2012dk}
Martin Holthausen, Manfred Lindner, and Michael~A. Schmidt.
\newblock {CP and Discrete Flavour Symmetries}.
\newblock {\em JHEP}, 04:122, 2013, 1211.6953.

\bibitem{Chen:2014tpa}
Mu-Chun Chen, Maximilian Fallbacher, K.~T. Mahanthappa, Michael Ratz, and
  Andreas Trautner.
\newblock {CP Violation from Finite Groups}.
\newblock {\em Nucl. Phys.}, B883:267--305, 2014, 1402.0507.

\bibitem{Harrison:2002er}
P.~F. Harrison, D.~H. Perkins, and W.~G. Scott.
\newblock {Tri-bimaximal mixing and the neutrino oscillation data}.
\newblock {\em Phys. Lett.}, B530:167, 2002, hep-ph/0202074.

\bibitem{Xing:2002sw}
Zhi-zhong Xing.
\newblock {Nearly tri bimaximal neutrino mixing and CP violation}.
\newblock {\em Phys. Lett.}, B533:85--93, 2002, hep-ph/0204049.

\bibitem{Esteban:2016qun}
Ivan Esteban, M.~C. Gonzalez-Garcia, Michele Maltoni, Ivan Martinez-Soler, and
  Thomas Schwetz.
\newblock {Updated fit to three neutrino mixing: exploring the
  accelerator-reactor complementarity}.
\newblock {\em JHEP}, 01:087, 2017, 1611.01514.

\bibitem{DellOro:2016tmg}
Stefano Dell'Oro, Simone Marcocci, Matteo Viel, and Francesco Vissani.
\newblock {Neutrinoless double beta decay: 2015 review}.
\newblock {\em Adv. High Energy Phys.}, 2016:2162659, 2016, 1601.07512.

\bibitem{PDG2018}
M.~Tanabashi et~al.
\newblock {Review of Particle Physics}.
\newblock {\em Phys. Rev.}, D98:030001, 2018.

\bibitem{Jarlskog:1985ht}
C.~Jarlskog.
\newblock {Commutator of the Quark Mass Matrices in the Standard Electroweak
  Model and a Measure of Maximal CP Violation}.
\newblock {\em Phys. Rev. Lett.}, 55:1039, 1985.

\bibitem{Escobar:2008vc}
J.~A. Escobar and Christoph Luhn.
\newblock {The Flavor Group $\Delta (6 \, n^2)$}.
\newblock {\em J. Math. Phys.}, 50:013524, 2009, 0809.0639.

\bibitem{Albright:2008rp}
Carl H.~Albright and Werner Rodejohann.
\newblock {Comparing Trimaximal Mixing and Its Variants with Deviations from Tri-bimaximal Mixing}.
\newblock {\em Eur. Phys. J.}, C62:599, 2009, 0812.0436.

\bibitem{Lam:2006wm}
C.~S.~Lam.
\newblock {Mass Independent Textures and Symmetry}.
\newblock {\em Phys. Rev.}, D74:113004, 2006, hep-ph/0611017.

\bibitem{Jenkins:2007ip}
Elizabeth~Ellen Jenkins and Aneesh~V. Manohar.
\newblock {Rephasing Invariants of Quark and Lepton Mixing Matrices}.
\newblock {\em Nucl. Phys.}, B792:187--205, 2008, 0706.4313.

\bibitem{Branco:1986gr}
G.~C. Branco, L.~Lavoura, and M.~N. Rebelo.
\newblock {Majorana Neutrinos and CP Violation in the Leptonic Sector}.
\newblock {\em Phys. Lett.}, B180:264--268, 1986.

\bibitem{Nieves:1987pp}
Jose~F. Nieves and Palash~B. Pal.
\newblock {Minimal Rephasing Invariant CP Violating Parameters With Dirac and
  Majorana Fermions}.
\newblock {\em Phys. Rev.}, D36:315, 1987.

\bibitem{delAguila:1995bk}
F.~del Aguila and M.~Zralek.
\newblock {CP violation in the lepton sector with Majorana neutrinos}.
\newblock {\em Nucl. Phys.}, B447:211--226, 1995, hep-ph/9504228.

\bibitem{Aghanim:2018eyx}
N.~Aghanim et~al.
\newblock {Planck 2018 results. VI. Cosmological parameters}.
\newblock 2018, 1807.06209.

\bibitem{Hagedorn:2016lva}
Claudia Hagedorn and Emiliano Molinaro.
\newblock {Flavor and CP symmetries for leptogenesis and $0\nu\beta\beta$
  decay}.
\newblock {\em Nucl. Phys.}, B919:404--469, 2017, 1602.04206.

\bibitem{Feruglio:2013hia}
Ferruccio Feruglio, Claudia Hagedorn, and Robert Ziegler.
\newblock {A realistic pattern of lepton mixing and masses from $S_4$ and CP}.
\newblock {\em Eur. Phys. J.}, C74:2753, 2014, 1303.7178.

\bibitem{model}
C.~Hagedorn and J.~K{\"o}nig.
\newblock {Lepton and Quark Masses and Mixing in a SUSY Model with $\Delta
  (384)$ and CP}.
\newblock 2018, 1811.09262.

\bibitem{GAP1}
GAP.
\newblock {GAP -- Groups, Algorithms, and Programming, Version 4.5.5}.
\newblock {\em http://www.gap-system.org}, 2012.

\bibitem{GAP2}
H.~U. Besche, B.~Eick, and E.~O'Brien.
\newblock {SmallGroups- library of all 'small' groups, GAP package, Version
  included in GAP 4.5.5}.
\newblock {\em http://www.gap-system.org/Packages/sgl.html}, 2002.

\end{thebibliography}
\end{document}